\documentclass[rnote]{aa} 
\usepackage{graphicx}
\usepackage{txfonts}
\usepackage{natbib}
\bibpunct{(}{)}{;}{a}{}{,} 

\newcommand{\new}[1]{{ #1}}
\newcommand{\newtwo}[1]{{ #1}}
\newcommand{\newthree}[1]{{ #1}}
\begin{document}
   \title{A 10-hour period revealed in optical spectra of the highly variable WN8 Wolf-Rayet star WR\,123\thanks{Based on observations obtained at the Canada-France-Hawaii Telescope (CFHT) which is operated by the National Research Concil of Canada, the Institut National des Sciences de l'Univers of the Centre National de Recherche Scientifique of France, and the University of Hawaii. Based also on observations made with ESO Telescopes at the La Silla Observatory, under programme ID 271.D-5025.}$^,$\,\thanks{Photometric date presented in Figure~1 are available in electronic form at the CDS via anonymous ftp to \tt{cdsarc.u-strasbg.fr} (130.79.128.5) or via \tt{http://cdsweb.u-strasbg.fr/cgi-bin/qcat?J/A+A/ }}}

\author{A.-N. Chen\'e\inst{\ref{inst1},}\inst{\ref{inst2},}\inst{\ref{inst3}}
      \and
      C. Foellmi\inst{\ref{inst4}}
      \and
      S.V. Marchenko\inst{\ref{inst5}}
      \and
      N. St-Louis\inst{\ref{inst6}}, A.~F.~J. Moffat\inst{\ref{inst6}}
      \and
      D. Ballereau\inst{\ref{inst7}}, J. Chauville\inst{\ref{inst7}}
      \and
      J. Zorec\inst{\ref{inst8}}
      \and
      C.~A. Poteet\inst{\ref{inst9}}
      }

\institute{Departamento de Astronom\'ia, Casilla 160-C, Universidad de Concepci\'on, Chile \email{achene@astro-udec.cl} \label{inst1}
    \and
    Departamento de F\'isica y Astronom\'ia, Facultad de Ciencias, Universidad de Valpara\'iso, Av. Gran Breta\~na 1111, Playa Ancha, Casilla 5030, Valpara\'iso, Chile \label{inst2}
    \and
Canadian Gemini Office, Herzberg Institute of Astrophysics, 5071, West Saanich Road, Victoria (BC), Canada V9E 2E7 \label{inst3}
        \and
        17 bd Agutte Sembat, 38000 Grenoble, France \email{cedric.foellmi@gmail.com} \label{inst4}
    \and
    Science Systems and Applications, Inc., 10210 Greenbelt Road, Suite 600, Lanham, MD 20706, USA \email{sergey.marchenko@ssaihq.com} \label{inst5}
    \and
    D\'{e}partement de Physique and CRAQ, Universit\'{e} de Montr\'{e}al, C.P. 6128, Succ. Centre-Ville, Montr\'{e}al, Qu\'{e}bec, Canada H3C 3J7 \email{stlouis@astro.umontreal.ca, moffat@astro.umontreal.ca} \label{inst6}
    \and
    GEPI, UMR 8111 du CNRS, Observatoire de Paris-Meudon, 92195, Meudon, France \email{dominique.ballereau@obspm.fr, jacques.chauville@obspm.fr} \label{inst7}
    \and  
    Institut d'Astrophysique de Paris, UMR 7095 CNRS-Universit\'e Pierre \& Marie Curie, 98bis, bd. Arago, 75014 Paris, France \email{zorec@iap.fr} \label{inst8}
    \and
    Ritter Observatory, Department of Physics \& Astronomy, The University of Toledo, Mail Stop \#113, Toledo, Ohio 43606-3390, USA \email{cpoteet@physics.utoledo.edu} \label{inst9}
    }

   \date{Received ...; accepted ...}
 
  \abstract
   {}
   {What is the origin of the large-amplitude variability in Wolf-Rayet WN8 stars in general and WR123 in particular? A dedicated spectroscopic campaign targets the ten-hour period previously found in the high-precision photometric data obtained by the MOST satellite.}
  {In June-August 2003 we obtained a series of high signal-to-noise, mid-resolution spectra from several sites in the $\lambda\lambda$~4000~--~6940 \AA \, domain.  We also followed the star with occasional broadband (Johnson V) photometry. The acquired spectroscopy allowed a detailed study of spectral variability on timescales from $\sim\,5$ minutes to months.}
  {We find that all observed spectral lines of a given chemical element tend to show similar variations and that there is a good correlation between the lines of different elements, without any significant time delays, save the strong $\newtwo{absorption}$ components of the He{\sc i} lines, which tend to vary differently from the emission parts. We find a single sustained periodicity, $P \sim 9.8$\,h, which is likely related to the relatively stable pulsations found in MOST photometry obtained one year later. In addition, seemingly stochastic, large-amplitude variations are also seen in all spectral lines on timescales of several hours to several days.}
   {}


   \maketitle

\section{Introduction}

Population I Wolf-Rayet (WR) stars of the WN8 subclass have been the subject of numerous studies owing to their ``enigmatic'' behavior. These direct descendants of massive and hot Of stars are markedly distinct from other WR stars. Indeed, no member of this class is known to have a close massive O-type companion. They avoid star clusters and associations. Their runaway frequency is relatively high, and in some cases their distance from the Galactic plane is large \citep{Mof80,Mof89,Mof98}. In addition, all photometric, polarimetric and spectroscopic monitoring campaigns of WN8 stars have consistently shown  the highest level of intrinsic variability among all WR stars \citep[e.g.;][]{Mof86,Lam87,Dri87,Ba89,Rob89,Rob92,Ant95,Mar98m}. However, despite the claims of multiple photometric periods ranging from hours to days, there is no period that could be labeled as unique to all the campaigns for any given WN8 star. Discovery of any stable periodicity pattern in a WN8 star  would help to establish whether the star is a binary or a single rotator with an inhomogeneous wind or, alternatively, to refine the rather controversial pulsation models of WR stars. 

All of the above applies specifically to our target, WR\,123 (see Marchenko et al. 1998 for references), which shows an extremely complex, rapidly evolving light curve, along with  seemingly aperiodic  polarimetric \citep{StL} and  high-amplitude line-profile variability \citep{Mas,Lam83,Mar98m}. All searches for periodic signals provided  rather inconsistent results until 2004, when an intense, long-term, broad-band photometry campaign of WR\,123 \citep{Lef} was performed by the {\it MOST (Microvariability and Oscillations of STars)} satellite. It was shown that large-scale, short-lived (typically, less than 1 week) photometric fluctuations were dominated by the ever-variable continuum. Only one relatively stable, small-amplitude period was present in the entire data set at $\rm \newthree{P \approx 9.8}$\newthree{\,h}. Does this period indicate the detection of a second spectroscopic binary system among WN8 stars\footnote{The runaway WR148 = HD197406 is the first detected WN8 spectroscopic binary with a moderate-mass, unseen companion \citep{Ma96}}  or does it provide a reliable observation of periodic stellar activity in a WR wind, like pulsations or co-rotating density structures?

While broadband photometry documents the variability of the (mainly) stellar continuum emanating from the inner WR wind, spectroscopy probes the layers outside it. Moreover, since the emission lines of different chemical species and/or different ionization potentials are generally formed in different regions of the wind, the timing and correlation of profile variations detected in such lines may help for determining how perturbations propagate in a stratified wind. In this paper we present the results of intense spectroscopic monitoring of WR\,123.

\section{Observations and data reduction}

Originally, we planned to obtain spectroscopic monitoring in parallel with the MOST run, initially scheduled for June/July 2003. Therefore, we carried out spectroscopic observations of WR\,123 during 5 runs on 4 different telescopes (Table~1). Unfortunately, the launch of MOST was delayed by one year.  A small-scale spectroscopic campaign run in parallel with the 2004 MOST observations proved to be of critical importance and provided a picture, though rather sketchy, of line-profile variability (lpv) superposed on an ever-changing continuum flux \citep{Lef}. Here, working with a unique, much more extensive data-set, we intend to obtain a more detailed description of the lpv, albeit outside the MOST campaign. The acquired spectroscopy allowed us to follow lpvs on timescales from $\sim\,5$ minutes (NTT spectra only) to months (all spectra).

\begin{table*}
   \centering
    \caption{Spectroscopic Observing Campaigns}
    \label{ParObsSp}
    \begin{tabular}{lllll}
      \hline
      & La Silla & Observatoire de & Canada-France- & Observatoire du \\
      & Observatory & Haute-Provence & Hawaii Telescope & mont M\'egantic \\
      \hline
      HJD ($-$ 2452800) & 15.7, 40.8 \new{--} 45.8 & 28.5 \new{--} 34.5 & 60.9 \new{--} 64.8 & 73.6\\
      Telescope & 3.58m (NTT) & 1.52m & 3.6m & 1.6m \\
      Instrument & EMMI & Aur\'elie & OSIS & OMM-Spectro \\
      No. of spectra & 290 & 66 & 17 & 2 \\
      Resolution, \AA & 5.1 (6 pix.) & 1.4 (3 pix.) & 1.5 (3 pix.) & 1.6 (3 pix.) \\
      Spectral range, \AA & 4000-6940  & 5225-6100  & 4400-6100  & 4500-6100  \\
      Signal-to-noise & $\ge$ 200 & 100 & 200 & 100 \\
      \hline
    \end{tabular}
\end{table*}

\new{In parallel with the 2003 spectroscopic monitoring, we also performed broadband (Johnson V)  photometry using the 24--inch telescope of the Bell Observatory (Western Kentucky University, USA). The differential photometry was carried out using the weighted combination of comparison stars in the $4.4\times6.6$ arcmin CCD field. We used comparisons with the following (epoch 2000) RA and DEC coordinates: \newtwo{($\alpha$ = 19:03:58.2; $\delta$ = \newthree{--}04:18:48),($\alpha$ = 19:04:04.6; $\delta$ = \newthree{--}04:17:35),  ($\alpha$ = 19:03:45.3; $\delta$ = \newthree{--}04:19:30), ($\alpha$ = 19:03:54.7; $\delta$ = \newthree{--}04:20:48)}, and 1.0:0.75:0.25:0.25 weights, as well as a star at \newtwo{($\alpha$ = 19:04:13.2; $\delta$ = \newthree{--}04:20:31)} as a control star.} We plot the available photometric data and corresponding $\pm \, 1 \, \sigma$ errors in Figure~1. All the spectroscopic and photometric data were processed using standard {\sc iraf}\footnote{{\sc iraf} is distributed by the National Optical Astronomy Observatory, which is operated by the Association of Universities for Research in Astronomy, Inc. under cooperative agreement with the National Science Foundation.} software packages. \new{Special care was taken for the normalization of the spectra. First, a mean was made for each run. Then each spectrum of a run was divided by the run mean and the ratio fitted with a low order Legendre polynomial (between 4$^{\rm th}$ and 8$^{\rm th}$ order). The original individual spectrum was divided by this fit, and was therefore at the same level as the run mean. When this procedure was done for each run, the run means were then put at the same level by using the same procedure as described above. This allowed us to put all individual spectra at the same level. Then, we combined all run means into a global, high quality mean, which was then fitted in selected pseudo-continuum regions, i.e. wavelength regions where large emission-lines do not dominate.  These regions \newtwo{are listed in Table~2}. To avoid having to use high order polynomials for the fit of a very large wavelength region, we cut the spectra in 5 different sections to fit, i.e. 3999 -- 4454\,\AA, 4400 -- 4794\,\AA, 4750 -- 5294\,\AA, 5229 -- 6032\,\AA\, and  5999 -- 7000\,\AA. Finally, the fitted continuum function is applied to each individual spectrum. The error on the continuum normalization measured as the standard deviation of individual spectra around the continuum function is typically of 0.5\,\%.}

\begin{table*}
   \centering
    \caption{\newtwo{Wavelength intervals of the selected pseudo-continuum regions}}
    \label{cont}
    \begin{tabular}{ccccc}
      \hline
            \newtwo{wav. int.}    &      \newtwo{wav. int.}     &      \newtwo{wav. int.}    &      \newtwo{wav. int.}     &      \newtwo{wav. int.}\\
              \newtwo{(\AA)}        &        \newtwo{(\AA)}        &        \newtwo{(\AA)}        &        \newtwo{(\AA)}         &        \newtwo{(\AA)}\\
      \hline
      \newtwo{4001 -- 4006} & \newtwo{4495 -- 4498} & \newtwo{5095 -- 5105} & \newtwo{5620 -- 5635} & \newtwo{6197 -- 6201}\\
      \newtwo{4044 -- 4047} & \newtwo{4585 -- 4589} & \newtwo{5245 -- 5255} & \newtwo{5760 -- 5766} & \newtwo{6335 -- 6347}\\
      \newtwo{4268 -- 4279} & \newtwo{4740 -- 4755} & \newtwo{5372 -- 5381} & \newtwo{5990 -- 5998} & \newtwo{6638 -- 6643}\\
      \newtwo{4420 -- 4430} & \newtwo{4830 -- 4840} & \newtwo{5515 -- 5525} &\newtwo{ 6021 -- 6025} & \newtwo{6771 -- 6794}\\
      \newtwo{4422 -- 4428} & \newtwo{4965 -- 4975} & \newtwo{5555 -- 5565} & \newtwo{6057 -- 6061} & \newtwo{6927 -- 6933}\\
      \hline
    \end{tabular}
\end{table*}

   \begin{figure}[htbp]
   \centering
   \includegraphics[width=.45\textwidth]{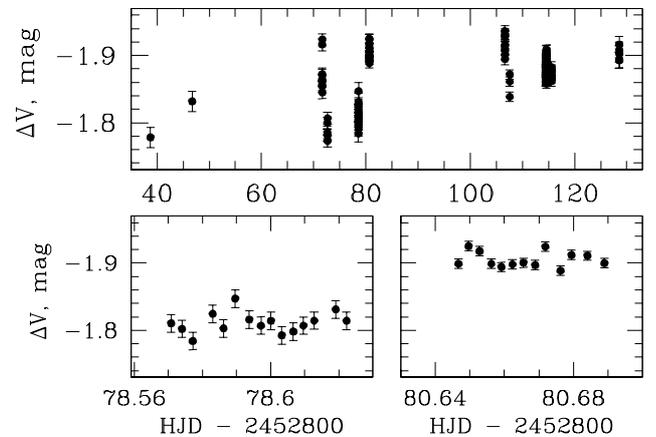}
      \caption{\new{The contemporaneous broad-band (Johnson V) photometry of WR\,123 obtained at the Bell Observatory. $\Delta$V is the  difference between the magnitudes of WR\,123 and comparison stars (see text). Each measurement is plotted with corresponding $\pm\, 1\,\sigma$ error bars. The upper panel shows all acquired data while the lower panel depicts two nights from August 2003.}}
         \label{photom}
   \end{figure}

\section{Results}

\subsection{Temporal variance spectrum}\label{VarLev}

In order to identify the variable parts of the spectrum of WR\,123, we adopt the ``temporal variance spectrum'' (TVS) approach developed by \citet{Full}. To minimize the influence of intrinsic radial velocity (RV) variations of WR\,123 on the TVS, we co-align the spectra using cross-correlation. \new{All the spectra were resampled after the co-alignment at the resolution of the spectra obtained at NTT, since they are the most numerous in our dataset and they have the lowest spectral resolution. The resampling was made using a sinc function. This method tends to slightly broaden the line profiles (negligibly in the case of WR spectral lines), but it does not modify the level of noise in the spectra and, therefore, does not affect the TVS.} We find that {\it all} the spectral lines in the observed spectral domain,  $\lambda\lambda \, 4000-6940$\,\AA , show lpvs significantly above the adopted 99\,\%  threshold. All reasonably isolated spectral lines produce a similar shape in the normalized TVS plot. However, in a given P~Cygni profile the absorption displays  a single narrow peak in the normalized TVS spectrum, and the emission part shows a wider and more structured peak with stronger variability in the central parts (e.g., Figure~2 with He{\sc ii} $\lambda$ 5412\,\AA\, and He{\sc i} $\lambda$ 5876\,\AA\, as typical examples).  This indicates that the detected lpv are not directly related to continuum variability. The latter would create a predictable global pattern with a shape following the emission-line contours.

\begin{figure*}
 \label{TVS} 
   \centering
   \includegraphics[width=.95\textwidth]{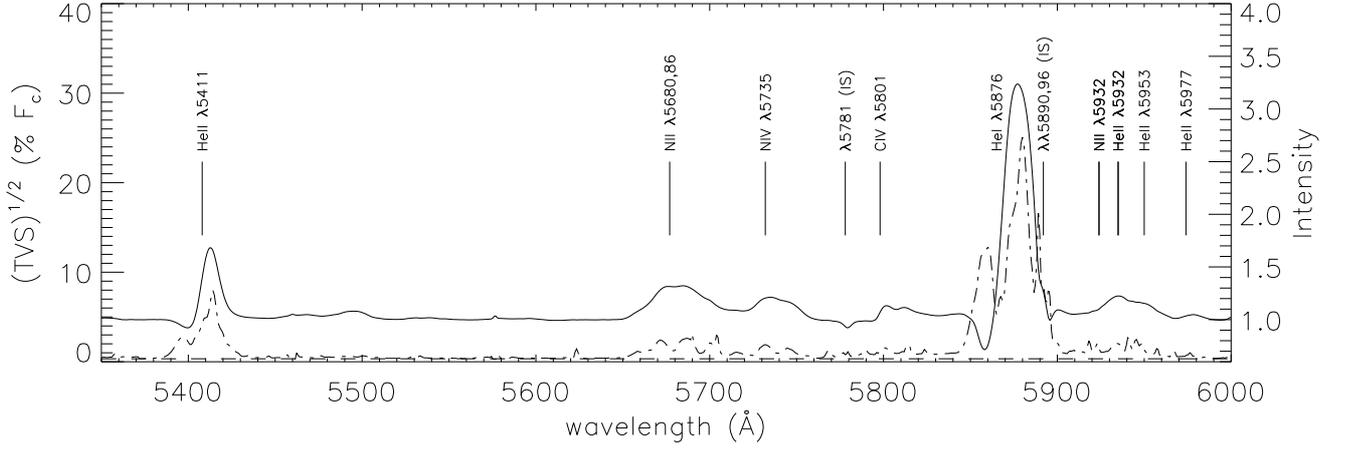}
 \caption{Mean rectified spectrum (solid line) of WR\,123 calculated using all the spectra from all the observing runs between 5350\,\AA\, and 6000\,\AA\, spectral range (common to all runs). The TVS spectrum (dash-dotted line) obtained using the formalism of \citet{Full} is overplotted. The dashed line is the threshold of significant variability at a level of 99\,\% for the TVS.}
\end{figure*}

\subsection{Line-profile and equivalent-width variability}\label{lpewv}

The TVS analysis shows that, in absolute terms, the most variable and [relatively] isolated lines are He{\sc i} $\lambda$ 4471, 4922, 5016, 5876, 6678\,\AA; He{\sc ii} $\lambda$ 4686, 4860, 5412, 6560\,\AA; and N{\sc iii}$ \lambda\lambda$ 4640,41\,\AA.  

To study the nightly changes in the line profiles, we single out two strong and least blended transitions, He{\sc ii} $\lambda$ 5412\,\AA\, and He{\sc i} $\lambda$ 5876\,\AA. The most variable part of the He{\sc i} $\lambda$ 5876\,\AA\, line is its absorption component, though the emission also changes from night to night. No apparent recurrence or coherent behavior in the variability patterns can be found at first glance.  The He{\sc ii}$\lambda$5412\AA\, line does not precisely follow the changes in He{\sc i} $\lambda$ 5876\,\AA. However, the lines behave in a roughly similar manner, i.e. the amplitudes of the absorption and emission components vary strongly and seemingly randomly on a timescale of days and longer.

From inspection of the remaining lines we conclude that what determines the global variability patterns of a line on a timescale of days is the strength of the P~Cygni absorption component. Indeed, regardless of their species, all lines with a relatively weak absorption component change similarly to He{\sc ii}  $\lambda$ 5412\,\AA\, while all lines with a strong P Cygni absorption  behave like He{\sc i}  $\lambda$ 5876\,\AA. 

On a timescale of hours all the lines show similar lpv in shape and in relative intensity, if one accounts for spectral blending effects. Within a given night, intensity and width can vary from 5\,\% to 30\,\%, with the latter, if applied to strong emission features, in  clear excess of the variability expected from the ever-changing continuum level \new{on the same timescale}. 

In an attempt to quantify the changes for each of our selected lines, we examine their equivalent widths (EW). Since the variability of the absorption and emission components of P~Cygni profiles do not appear to be correlated, we have treated their EW measurements separately by fitting two gaussian functions, one with a
negative intensity and displaced from the line center (absorption) and another with a positive intensity (emission).  The measurement errors \new{were} obtained following \citet{Vol}. Though realizing that such an approach may not be applicable to all considered cases, we note that most of the 2-component fits have a reasonably low $\chi^2$
and produce fairly good results for the blue-side portions of profiles.\\
In an attempt to sample a significant part of the WR wind we choose the He{\sc i} $\lambda$ 4471\,\AA\, (which gives better fits than He{\sc i} $\lambda$ 5876\,\AA) and He{\sc ii} $\lambda$ 5412\,\AA\, lines, both of which are strong, isolated and relatively easy to fit. We find that, in general, a stronger emission corresponds to a stronger absorption, and {\it vice versa}. However, for the He{\sc ii} $\lambda$ 5412\,\AA\, line, there are frequent disparities between the variability patterns of the emission and absorption components. Moreover, even if both lines (He{\sc i} and He{\sc ii}) vary with comparable amplitudes on a timescale of days, their variability within one night can be completely different. Taking into consideration the rest of the measured lines, we find that the EW of the emission parts of  the He{\sc ii}, N{\sc iii} and N{\sc iv} profiles are well-correlated, while quite frequently the He{\sc i} lines lack correlation with any other species and among themselves. This could be partially blamed on the presence of strong absorption components with a pronounced tendency to show different lpv patterns. The lack of correlation is especially prominent in the long-term (days-months). However, a tight He{\sc i} - He{\sc ii} correlation may reappear in some of the short-term (nightly) data segments.

\begin{figure*}[htbp]
 \label{periods} 
   \centering
   \includegraphics[width=.95\textwidth]{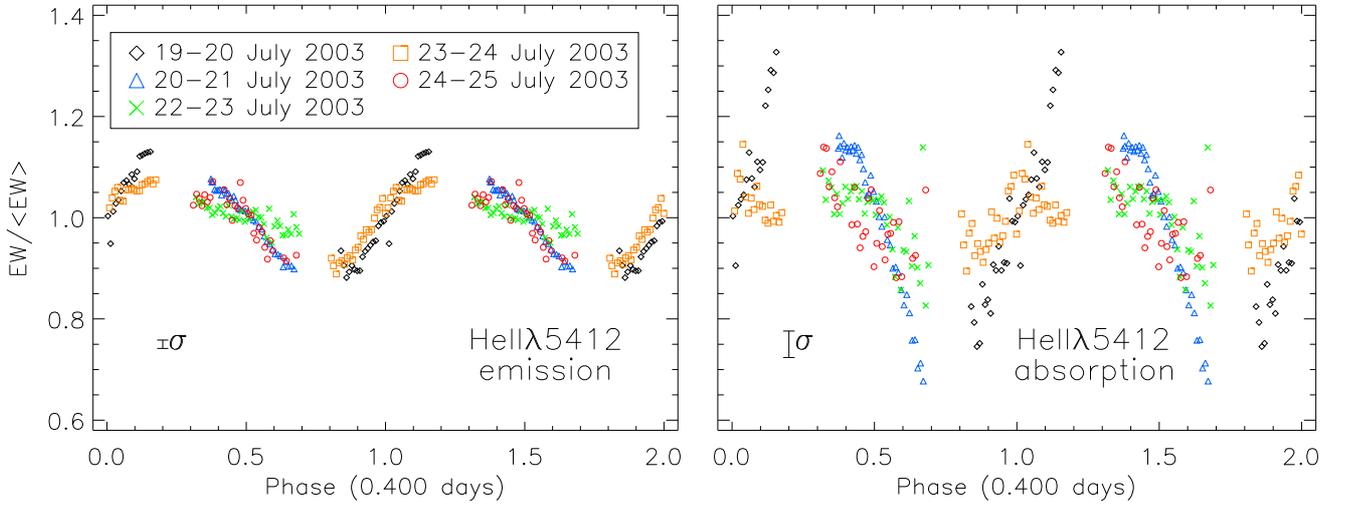}
 \caption{The de-trended EW measurements of the emission component ({\it left panel}) and of the absorption component ({\it right panel}) of the He{\sc ii} $\lambda$ 5412\,\AA\, line folded in phase with $\rm P\,=\,\newthree{0\fd400}$ and an arbitrary 0 phase. Five half-nights of the second NTT run are shown along with the corresponding $1\,\sigma$ error bars.}
\end{figure*}

\subsection{Radial velocity and skewness variability}\label{RV_sk}

Previous RV studies of WR 123 have consistently shown the presence of small-amplitude velocity  variations with $K\,\le\, 25$~km~s$^{-1}$ \citep{Mar98m}. Unfortunately, the NTT and CFHT data produced large systematic errors related to specific instrumental setups. This leaves OHP as the only reliable source of RV. We aligned the OHP spectra via cross-correlation with the highest S/N spectrum (the first spectrum of the campaign in this case), thus producing a mean OHP spectrum.  Then we used this mean to perform a cross-correlation with individual spectra one more time \new{and measure their RV. RVs were checked with the bisectors of the strongest emission lines. While the bisectors were giving bigger errors, they match the results coming from cross-correlation and show that all the lines are shifting in time with the same fashion.}

On a weekly time-scale, the measured \newtwo{RVs} vary by up to $\pm\,20$~km~s$^{-1}$, well in excess of the typical errors, $\sigma \,\sim \,2-5$~km~s$^{-1}$ \new{(obtained from the fit of the cross-correlation profile)}, by gradually decreasing from $\sim\,0$~km~s$^{-1}$ to $\sim\,-40$~km~s$^{-1}$ (with respect to the first spectrum of the campaign). A closer look shows that the RVs vary from night to night and can change by as much as 10~km~s$^{-1}$ within a $\sim\,4$-hour interval. Are these RV variations caused by the presence of a companion or simply by large-scale lpv?  One \newtwo{possible} way to verify this is to examine the skewness of the main spectral lines, i.e. He{\sc ii} $\lambda$ 5412\,\AA\, and He{\sc i} $\lambda$ 5876\,\AA. Since these two lines clearly dominate the spectrum (and therefore the flux in the cross-correlation profile) in the chosen wavelength range, they are also the main contributors to the RV measurements. We find that the skewness varies on a short timescale of $\sim$~hours. Hence, the variability of the RVs may be induced by large-scale, short-term profile variability, save the long-term downslope in the RV curve. It is not possible to relate any particular change in the overall line shape to a particular change in RV. With some caution, we conclude that the only ``genuine'', i.e., caused by a possible spatial (orbital?) motion of the star, RV variability may occur on a fairly long timescale of days or weeks.

\new{
\subsection{Photometric variations}

Our light curve (Figure~1) has fairly similar characteristics to the light curve published by \citet{Lef}. Both have a variation amplitude of $\sim$\,15--20\,\% on the long term and $\sim$\,5\,\% on the short term. Hence, we can assume that nothing major in the star and the star's wind has changed between the 2003 \newtwo{spectroscopic} campaign (this study) and the 2004 MOST campaign \citep{Lef} and that the two datasets can be compared. However, the limited photometric accuracy and the poor time sampling of our data compared with the unique MOST observations does not allow us to say anything more about WR\,123's photometric variability.
}

\section{Period search}

We performed a period search in the permittable range from 10 min up to one month, using two complementary approaches. The phase-dispersion minimization algorithm \citep{pdm} (PDM) is well-suited for cases with only a small number of observations and highly non-sinusoidal underlying signal. We also used the {\sc clean} algorithm \citep{Scargle,clean}, which has the advantage of taking into account the unevenness of the data sampling.

We searched for periodic signals in the RV data and found that the OHP subset reveals no significant periodicities.  Period search on the nightly mean EWs of the He{\sc ii} $\lambda$ 5412\,\AA\, line provides inconclusive results, mainly due to the scarcity of these particular (1 point/night) data. Searching for short periods, we concentrated on our numerous NTT spectra, first carefully removing any long-term (days-weeks) trends. \new{We find a period of $\newthree{0\fd400} \,\pm\, \newthree{0\fd004}$ ($\nu\,=\,2.50\,\pm\, 0.06\, \rm\, d^{-1}$), which is compatible with the range of frequencies between $2.40 \rm \,d^{-1}$ and $2.50 \, \rm d^{-1}$}  claimed by \citet{Lef} from MOST data\newtwo{,} is present in the de-trended EW measurements of both He{\sc i} (peak at $2.46\rm\, d^{-1}$) and He{\sc ii} lines (peak at $2.51\rm \,d^{-1}$) and is clearly detected by PDM, but only marginally present in the {\sc clean}ed PS. The detrended EW values of He{\sc ii} $\lambda$ 5412\,\AA\, folded in phase with this period are shown in Figure~3. 

\new{Since the amplitude of the photometric variations on the timescale of a few hours is relatively small compared with the accuracy of our photometric dataset and \newtwo{due} to our poor time sampling, no clear period could be found in the light curve.}

\section{Discussion}\label{Discussion}

Most of the lines in the WR~123 spectrum have a P~Cygni profile, i.e. they contain a blue-shifted absorption component formed in the column of wind between the stellar core and the observer, as well as an emission component formed in a \new{(more or less)} symmetric shell surrounding the core. Since the two components produce different variability patterns, we treated them separately wherever feasible.

All the emission components vary with similar shapes and relative amplitudes, up to 30\,\% (peak-to-valley) of an average line intensity on a short timescale (hours), and up to 55\,\% on a long timescale (days). There is no delay between the variability patterns of the different lines; i.e., the variability seems to occur simultaneously in the different line-forming regions of  the wind. 

The absorptions have a seemingly higher level of variability, up to 50\,\% of their depth on a short timescale, and as high as 90\,\% on a long timescale. However, this difference may come from the intrinsic difference between the formation process and the different range of intensity of the absorption (1 to 0 relative intensity) and the emission (1 to I relative intensity, where I is the intensity at the peak) parts of P~Cygni profiles. Moreover, the variations of the two parts of any given P~Cygni profile have similar shapes, indicating that they can be caused by the same phenomenon. All absorption components vary in concert, without detectable delay between different line species. \new{Considering that the star shows a consistent level of photometric variability that practically never exceeds $\sim\,0.15$ mag long-term \citep[days-month: e.g. ][\newtwo{our study}]{Mar98m,Lef} and rarely exceeds $\sim\,0.05$ mag during any particular night of ground-based observations (e.g., Figure~1), such a level of lpv cannot be a consequence of the continuum variations alone.} We note that for WR 123 the continuum flux is the dominating (usually up to 80\,\%) component in the broadband photometric optical flux. 

In the following we concentrate on the \newthree{0\fd4} period found in the de-trended EW data. We plot the period-folded lpv of the strongest and least blended lines He {\sc ii} $\lambda$ 5412\,\AA, He{\sc i} $\lambda$ 5876\,\AA\, and He{\sc ii} $\lambda$ 6560\,\AA\, in Figure~4. In the bottom parts of the panels we plot average profiles derived from four different phase intervals, $\phi\,=\,0.0-0.2$, $\phi\,=\,0.3-0.5$, $\phi\,=\,0.5-0.7$, $\phi\,=\,0.8-1.0$. It is apparent that practically the whole wind responds to the periodic perturbation coming from the core. In general, the absorption components are more involved in the periodic changes, as they come from a more localized region of the wind compared to the emissions. This does not hold true for He{\sc i}$\lambda$5876\AA\, which has the largest line-forming region among the considered lines, thus the best ability to ``smooth'' out the phase-dependent variability patterns.

\begin{figure}[htbp]
 \label{grisperiod1}
   \centering
   \includegraphics[width=.45\textwidth]{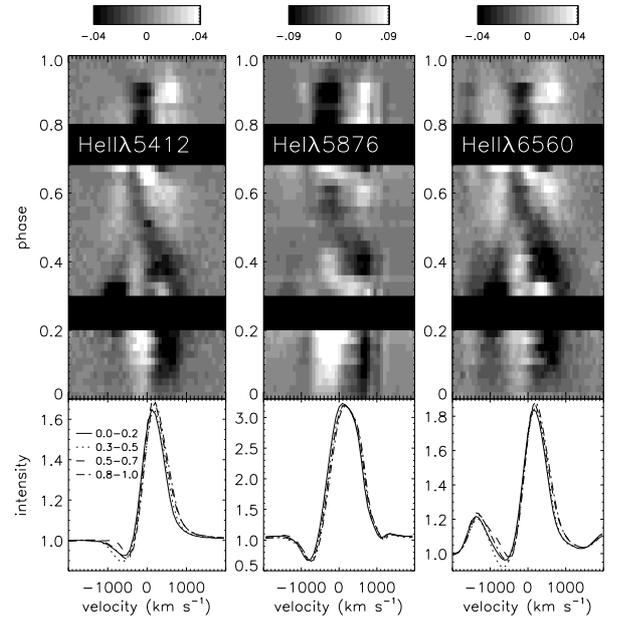}
 \caption{{\it Upper}:~Lpvs observed in the He{\sc ii} $\lambda$ 5412\,\AA\, (left),  He{\sc i}$ \lambda$ 5876\,\AA\, (center) and He{\sc ii} $\lambda$ 6560\,\AA\, (right) folded with the $\newthree{0\fd4}$-period. Long-term trends have been removed from the spectra and the remainders were enhanced by a factor of 2 for clarity. {\it Lower}:~Binned profiles over the four phase intervals $\phi\,=\,0.0-0.2$, $\phi\,=\,0.3-0.5$, $\phi\,=\,0.5-0.7$, $\phi\,=\,0.8-1.0$.}
\end{figure}

\citet{Lef} already dismissed the possibility that stellar rotation could be at the origin of the 9.8 hour period, since this would lead to a rotational velocity greater than the breakup speed at the surface of a typical WN8 star  \citep[when taking $R_*\,\sim\,15\, R_\odot$ according to ][]{Crowther}. Also, they showed that such a short period could not be produced by the presence of a (compact) companion revolving around the star, since the deduced orbit would be smaller than the stellar hydrostatic radius of WR\,123. Finally, they placed a firm upper limit, $A\,<\,0.2 \rm \,mmag$ for $\rm P \,\la \,1$ hour, on the amplitudes of strange-mode pulsations predicted by \citet{Glatzel}.

Since the publication of the MOST light curve, two independent models involving non-radial pulsations were proposed to explain the 9.8 hour period.  The first, by \citet{Tow}, examines the stability of $\ell\,=\,1$ and $\ell\,=\,2$ g-modes and finds that the $\kappa$-mechanism associated with a hot Fe-opacity bump within the star can be responsible for periods ranging between 3 and 12 hours. The second model proposed by \citet{Dor} contradicts the conclusion of \citet{Lef} concerning strange-mode pulsations. Indeed, using an adequate stellar radius for WR\,123, they showed that strange-modes with a period as high as $\sim\,10$ hours can be excited. However, both models exploited stellar parameters beyond the usually adopted range: \citet{Tow} used a stellar radius that is far too small for a WN8 star, and \citet{Dor} used a hydrogen abundance which is too high for WR\,123. More recently, \citet{Gla} followed the evolution of strange mode instabilities into the non-linear regime and showed that they generate consecutive shock waves which inflate the stellar envelope considerably and thus increase the pulsation periods, giving final periods that are consistent with that observed in WR 123. It remains unclear how these pulsations will affect the stellar wind (thus the wind-formed spectral lines) at different radii. Our data show that the 9.8-hour spectral variability involves all prominent lines in the observed spectral region, thus {\it globally} affecting the wind. 

Finally, on a speculative note inspired by the suggestion of \citet{Foe}, one might wonder if a Thorne-$\dot{\textrm{Z}}$ytkow-like scenario could be operating in WR123 and other WN8 stars and exciting stellar pulsations. Thorne-$\dot{\textrm{Z}}$ytkow Objects \citep[T$\dot{\textrm{Z}}$O,][]{Tho77} are assumed to be red supergiants with a degenerate neutron core. They are formed via merger of a massive star and a remnant of a supernova explosion in a close binary system.  Interestingly, a T$\dot{\textrm{Z}}$O nature of WN8 stars would explain their relatively large stellar radii, high large space velocities, the avoidance of OB associations and the (current) dearth of binaries.

\section{Conclusions}

Summarizing the results of our intense, multi-site monitoring of the variable 
WR star WR 123 in 2003, we note that: 
   \begin{enumerate}
      \item the star shows pronounced long-term, large-scale, seemingly stochastic spectral and photometric 
            variability, broadly in line with all previous studies;  
      \item the only stable period which is clearly present in the EW of prominent spectral lines matches the
            P\,=\,9.8-hour period found in the MOST photometry obtained one year later;            
      \item practically all the observed wind volume is involved in the 10-h periodic variations.
            Though the variability patterns may differ among the lines with different ionization potentials, 
            all the lines vary without measurable phase delays. This points to the stellar core as  
            sole driver of the periodic variability. 

   \end{enumerate}

\begin{acknowledgements}
ANC gratefully acknowledges support from the Chilean {\sl Centro de Astrof\'\i sica} FONDAP No. 15010003 and the Chilean Centro de Excelencia en Astrof\'\i sica y Tecnolog\'\i as Afines (CATA).  NSL and AFJM wish to thank the Natural Sciences and Engineering Research Council (NSERC) of Canada for financial support. 
\end{acknowledgements}

\bibliographystyle{aa} 
\bibliography{ms_ref.bib} 

\end{document}